\documentstyle[prl,aps,twocolumn,mypsfig]{revtex}

\begin{document}



\title{Approximate quantum error correction can lead to better codes}

\author{Debbie W. Leung$^{1}$, M. A. Nielsen$^{2,3}$, Isaac
	L. Chuang$^{2,4}$, {\em and}\/ Yoshihisa Yamamoto$^{1}$}

\address{\vspace*{1.2ex}
        \hspace*{0.5ex}{$^1$ ERATO Quantum Fluctuation Project \\
        Edward L. Ginzton Laboratory, Stanford University,
                Stanford, CA 94305-4085}}
\address{\vspace*{1.2ex}
        \hspace*{0.5ex}{$^2$ Institute for Theoretical Physics \\
        University of California, Santa Barbara, CA 93106-4030 }}
\address{\vspace*{1.2ex}
        \hspace*{0.5ex}{$^3$ Center for Advanced Studies, Department
        of Physics and Astronomy, \\
        University of New Mexico, NM 87131-1156}}
\address{\vspace*{1.2ex}
	\hspace*{0.5ex}{$^4$Theoretical Astrophysics T-6 \\
	Los Alamos National Laboratory, Los Alamos, NM 87545}}

\date{\today}
\maketitle


\begin{abstract}
We present relaxed criteria for quantum error correction which are
useful when the specific dominant quantum noise process is known.
These criteria have no classical analogue.  As an example, we provide
a four-bit code which corrects for a single amplitude damping error.
This code violates the usual Hamming bound calculated for a Pauli
description of the error process, and does not fit into the GF(4)
classification.
\end{abstract}



\section{Introduction}

Quantum error correction is the reversal of part of the adverse
changes due to an undesired and unavoidable quantum process.  Code
criteria for perfect error correction have been
developed\cite{Ekert96b,Knill96,Bennett96a,Nielsen96c}, and all
presently known codes satisfy these criteria exactly.  Furthermore,
most of these codes belong to a general GF(4)
classification\cite{Calderbank96a}.  In these schemes, quantum bit
(qubit) errors are described using a Pauli ($I$, $X$, $Y$, $Z$) error
basis, and coding is performed to allow correction of arbitrary
unknown errors.

However, in the usual case in the laboratory, one works with a specific
apparatus with a particular dominant quantum noise process.  The GF(4) codes
do not take advantage of this specific knowledge, and may thus be sub-optimal,
in terms of transmission rate and code complexity.  Unfortunately, there is no
general method to construct quantum codes in a non-Pauli basis, and few such
codes are known\cite{Chuang96a,Braunstein96a,Cirac96a,Plenio96a,Chuang96c}.
The code criteria are generally
very difficult to satisfy, and without the Pauli basis, no way is known to
apply classical coding techniques for quantum error correction.

In this paper, we develop a novel approach to quantum error correction
based on approximate satisfaction of the existing quantum error correction
criteria.  These approximate criteria are simpler and less restrictive, 
and in certain cases, 
such as for amplitude damping, codes can be found
relatively more easily.  The approximate criteria also admit more codes, 
therefore, codes requiring shorter block lengths may also be possible.

For example, using this approach we have discovered a four bit code
which corrects for single qubit {\em amplitude
damping}\cite{Chuang96c,Chuang96b} errors.  Such a short
{\em non-degenerate} code is impossible using the Pauli basis.  The reason
is that the effects \cite{Kraus83a,Schumacher96a} to be
corrected in a non-degenerate code have to map the codeword space to
orthogonal spaces if the syndrome is to be detected unambiguously.
Hence, the minimum allowable size for the encoding space is the product of
the dimension of the codeword space and the number of effects to be
corrected. The single qubit amplitude damping effect operators expressed 
in the Pauli basis are:
\begin{eqnarray}
        A_0 &=& \frac{1}{2}\left[\rule{0pt}{2.4ex}
                (1+\sqrt{1-\gamma}) I + (1-\sqrt{1-\gamma}) \sigma_z
                \right]
\label{eq:adzero}
\\      A_1 &=& \frac{\sqrt{\gamma}}{2} \left[\rule{0pt}{2.4ex}{ \sigma_x
                +
                i \sigma_y }\right]
\label{eq:adone}
\,.
\end{eqnarray}
These describe the changes due to the loss of zero or one excitation to
the environment.  To first order in the scattering probability
$\gamma$, $n+1$ possible effects may happen to an $n$-qubit code using
the $A_0$, $A_1$ error basis, so it follows that $n\geq 3$
is required. In contrast, in the Pauli basis, any $\sigma_x$ or
$\sigma_y$ error must be corrected by the code, so that $2n+1$
possible effects must be dealt with.  It follows that at least $n\geq
5$ is required for a non-degenerate Pauli basis code, in contrast to
the four bit code which we demonstrate in this paper.

The lessons are that (1) better codes may be found for specific error
processes, and (2) approximate error correction simplifies code construction 
and admits more codes.  Approximate error correction is a property
with no classical analogue, because it makes use of slight
non-orthogonalities possible only between quantum states.  We describe
our approach to this problem by first exhibiting our four-bit example
code in detail.  We then generalize our results to provide new,
relaxed error correction criteria and specific procedures for
decoding and recovery.  We conclude by discussing possible extensions
to our work.

\section{Four Bit Amplitude Damping Code}

\label{sec:noisemodel}

\begin{figure}[htbp]
\begin{center}
\mbox{\psfig{file=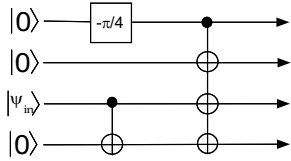,width=1.5in}}
\end{center}
\caption{Circuit for encoding a qubit.  The third mode contains the input
         qubit.  The rotation gate in the first qubit performs 
         $\exp(-i \pi \sigma_y/4 )$.}
\label{fig:enc}
\end{figure}

Consider the single qubit quantum noise process defined by
\begin{eqnarray}
        {\cal E}(\rho)&=&\sum_{k=0,1} A_k \rho A_k^{\dagger}
\label{eq:opsumrep}
\\
	A_0 & = & \left[ \begin{array}{cc}
	{1}&{0}\\{0}&\sqrt{1-\gamma} \end{array} \right] ~~~~~ 
        A_1 = \left[ \begin{array}{cc}
	{0}&{\sqrt{\gamma}}\\{0}&{0} \end{array} \right] 
\label{eq:ampdamperr}
\,, 
\end{eqnarray} 
known as {\em amplitude damping}.
The probability of losing a photon, $\gamma$, is assumed to be
small. To correct errors induced by this process, we encode one qubit
using four, with the logical states 
\begin{eqnarray} 
	|0_{L}\rangle &=&\frac{1}{\sqrt{2}}
	\left[\rule{0pt}{2.4ex}{ |0000\rangle + |1111\rangle }\right] 
\\ 	
	|1_{L}\rangle &=&\frac{1}{\sqrt{2}}
	\left[\rule{0pt}{2.4ex}{ |0011\rangle + |1100\rangle }\right]
\label{code}
\,.
\end{eqnarray}
A circuit for encoding the logical state is shown in Fig.\ref{fig:enc}.

The possible outcomes after amplitude damping may be written as
\begin{equation}
        [\psi_{out}\rangle  = \bigoplus_{\tilde{k}} |\phi_{\tilde{k}}\rangle 
        \equiv \bigoplus_{\tilde{k}} A_{\tilde{k}} |\psi_{in}\rangle  
\,, 
\end{equation}
where $\tilde{k}$ are strings of $0$'s and $1$'s serving as indices
for each error, and we use the notation $A_{010\cdots} = A_0 A_1
A_0\cdots$. $[\cdot\rangle $ is convenient shorthand\cite{Chuang96b} for a
mixed state (tensor sum of un-normalized pure states).  In the
following, the squares of the norm of $|\cdot\rangle $ states 
will give their probabilities
for occurring in a mixture.  For the input qubit state 
\begin{equation} 
	|\psi_{in}\rangle  = a |0_L\rangle  + b |1_L\rangle 
\label{eq:initial}
\,,
\end{equation}
{\em all} possible final states occurring with probabilities ${\cal
O}(\gamma)$ or above are
\begin{eqnarray}
\label{eq:decayfirst}
        |\phi_{0000}\rangle 
        &=& a\left[\rule{0pt}{2.4ex}{
	\frac{|0000\rangle +(1-\gamma)^{2}|1111\rangle }{\sqrt{2}}}\right]
\\
\nonumber
        &+& b\left[\rule{0pt}{2.4ex}{
	\frac{(1-\gamma)[|0011\rangle +|1100\rangle ]}{\sqrt{2}}}\right]
\\
\nonumber
        |\phi_{1000}\rangle 
        &=& \sqrt{\frac{\gamma (1-\gamma)}{2}} \left[\rule{0pt}{2.4ex}{
        a(1-\gamma)|0111\rangle  + b |0100\rangle  }\right]
\\
\nonumber
        |\phi_{0100}\rangle 
        &=& \sqrt{\frac{\gamma (1-\gamma)}{2}} \left[\rule{0pt}{2.4ex}{
        a(1-\gamma)|1011\rangle  + b |1000\rangle  }\right]
\\
\nonumber
        |\phi_{0010}\rangle 
        &=& \sqrt{\frac{\gamma (1-\gamma)}{2}} \left[\rule{0pt}{2.4ex}{
        a(1-\gamma)|1101\rangle  + b |0001\rangle  }\right]
\\
\nonumber
        |\phi_{0001}\rangle 
        &=& \sqrt{\frac{\gamma (1-\gamma)}{2}} \left[\rule{0pt}{2.4ex}{
        a(1-\gamma)|1110\rangle  + b |0010\rangle  }\right]
\,.
\end{eqnarray}
The usual criteria used to study quantum codes (reviewed in section 
\ref{sec:approxcond}) require that 
$\langle 0_L|A_{\tilde{k}}^\dagger A_{\tilde{k}}|0_L\rangle =
\langle 1_L|A_{\tilde{k}}^\dagger A_{\tilde{k}}|1_L\rangle $, but
\begin{eqnarray}
        \langle 0_L|A_{0000}^\dagger A_{0000}|0_L\rangle &=&
        1-2\gamma+3\gamma^2+{\cal O}(\gamma^3)
\\
        \langle 1_L|A_{0000}^\dagger A_{0000}|1_L\rangle &=&
	1-2\gamma+\gamma^2
\,.
\end{eqnarray}
So the code we have constructed does not satisfy the usual criteria.  
We will demonstrate that the code satisfies new approximate error correction
conditions later on, and revisit the recovery procedure afterwards.
First, we exhibit how it works.

\subsection{Decoding and Recovery Circuit}
\label{sec:reco}  
\begin{figure}[htbp]
\begin{center}
\mbox{\psfig{file=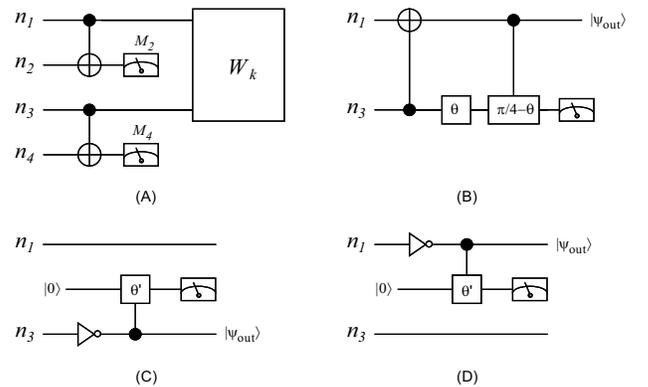,width=3.3in}}
\end{center}
\caption{(A) Circuit for error syndrome detection.  The measurement result
	is used to select $W_k$ out of three actions.  If the result 
	($M_2$, $M_4$)~is 00, 10, or 01,
	circuits (B), (C), or (D) are applied, respectively, to recover the
	state.  The angles $\theta$, $\theta'$ are given by 
	$\tan \theta = (1-\gamma)^2$ and $\cos \theta' = 1 - \gamma$. 
	The rotation gate and controlled-rotation gate specified 
	by the angle $\tilde{\theta}$
	perform the functions $\exp(i\tilde{\theta}\sigma_y)$ and 
	$\Lambda_1(\exp(i\tilde{\theta}\sigma_y))$ respectively
	in the notation of \protect\cite{Barenco95}.}
\label{fig:fbitall}
\end{figure}

Let us denote each of the four qubits by 
$|n_1\rangle$,$\ldots$,$|n_4\rangle$.  
Error correction is performed by distinguishing the five possible
outcomes of Eq.(\ref{eq:decayfirst}), and then applying the
appropriate correction procedure.  The first step is {\em syndrome
calculation}, which may be done using the circuit shown in
Fig.~\ref{fig:fbitall}A.  There are three possible measurement results
from the two meters: ($M_2$, $M_4$) = (0,0), (1,0) and (0,1).
Conditioned on ($M_2$, $M_4$), recovery processes $W_k$ implemented by the
other three circuits of Fig.~\ref{fig:fbitall} can be applied to the
output $|n_1n_3\rangle $ from the syndrome calculation circuit.

If ($M_2$, $M_4$) = (0,0), then $|n_1n_3\rangle $ is:
\begin{equation}
	a \left[\rule{0pt}{2.4ex}{ 
	\frac{|00\rangle  + (1-\gamma)^2 |11\rangle }{\sqrt{2}} }\right]
        + b \left[\rule{0pt}{2.4ex}{ 
	\frac{(1-\gamma) (|01\rangle  + |10\rangle )}{\sqrt{2}} }\right]
\,.
\end{equation}
To regenerate the original qubit, the circuit of Fig.~\ref{fig:fbitall}B
is used: a controlled-not is applied using $|n_3\rangle$ as the control, 
giving
\begin{equation}
	a |0\rangle  \left[\rule{0pt}{2.4ex}{ 
	\frac{|0\rangle  + (1-\gamma)^2 |1\rangle }{\sqrt{2}} }\right]
        +b |1\rangle  \left[\rule{0pt}{2.4ex}{ 
	\frac{(1-\gamma) (|1\rangle  + |0\rangle )}{\sqrt{2}} }\right]
\,.
\end{equation}
$|n_1\rangle $ can now be used as a control to rotate $|n_3\rangle $
to be parallel to $|0\rangle $. We obtain as the final output:
\begin{eqnarray}
        |n_1n_3\rangle  & = & \left[\rule{0pt}{2.4ex}{ 
	a \sqrt{\frac{(1-\gamma)^4+1}{2}}|0\rangle 
        + b(1-\gamma) |1\rangle  }\right] |0\rangle 
\\
      	&=& \left[\rule{0pt}{2.4ex}{
	(1-\gamma) (a|0\rangle + b|1\rangle) + {\cal O}(\gamma^2)|0\rangle 
	}\right] |0\rangle 
\label{eq:outcome1}
\,,
\end{eqnarray}
with the corrected and decoded qubit left in $|n_1\rangle $ as desired.
 
If ($M_2$, $M_4$) = (1,0), the inferred state before syndrome measurement 
is $\phi_{1000}$ or $\phi_{0100}$.  In either case, $|n_1 n_3 \rangle $ 
is in a product 
state and the third qubit has the distorted state:
\begin{equation}
        |n_3\rangle  = 
	\sqrt{\frac{(1-\gamma)\gamma}{2}} \left[\rule{0pt}{2.4ex}{
	a (1-\gamma)|1\rangle  + b |0\rangle  }\right]
\label{eq:dist}
\,.
\end{equation}
To undo the distortion, we apply the {\em non-unitary} transformation
in Fig.~\ref{fig:fbitall}C.  The combined operation on 
$|n_3 \rangle \langle n_3|$ due to the NOT gate, 
the controlled-rotation gate and the measurement of the ancilla bit can be 
expressed in the operator sum representation:
${\cal N}(\rho) = N_0 \rho N_0^\dagger + N_1 \rho N_1^{\dagger}$, where
\begin{equation}
        N_0 = |0\rangle \langle 1|+(1-\gamma)|1\rangle \langle 0|,
	~N_1  = \sqrt{\gamma (2 - \gamma)} |1\rangle \langle 0|
\,.
\end{equation}
The $N_0$ and $N_1$ operators correspond to measuring the ancilla
to be in the $|0\rangle$ and $|1\rangle$ states respectively.  
If the ancilla state is $|0\rangle$, we obtain the state:
\begin{equation}
        |\psi_{out}\rangle  = \sqrt{\frac{(1-\gamma)^3\gamma}{2}}
        \left[\rule{0pt}{2.4ex}{a |0\rangle  + b |1\rangle  }\right]
\,,
\end{equation}
in the third mode because $N_0$ preferentially damps out the 
$b|0\rangle$ component in Eq.(\ref{eq:dist}).
We get an error message if the ancilla is in the $|1\rangle$ state.  
Finally, if ($M_2$, $M_4$) = (0,1) the same procedure can be
applied as in the ($M_2$, $M_4$) = (1,0) case, with $n_1$ and $n_3$
swapped.

The fidelity, defined as the worst (over all input states) possible
overlap between the original qubit and the recovered qubit is
\begin{eqnarray}
        {\cal F} &=& (1-\gamma)^2 + 4 \left[\rule{0pt}{2.4ex}{
	\frac{(1-\gamma)^3 \gamma}{2} }\right]
\\      
        &=& 1 - 5\gamma^2 + {\cal O}(\gamma^3)
\,,
\end{eqnarray}
Note that the final state Eq.(\ref{eq:outcome1}) is slightly distorted.
This occurs because the recovery operation is not exact, due to the 
failure to satisfy the code criteria exactly.  Furthermore, the circuits in
Fig.~\ref{fig:fbitall}C and~\ref{fig:fbitall}D have a finite probability
for failure.  However, these are second order problems, and do not
detract from the desired fidelity order.

\section{Approximate Sufficient Conditions}
\label{sec:approxcond}
We now explain why our code works despite its violation of the usual
error correction criteria.  The reason is simple: small deviations
from the criteria are allowed as long as they do not detract from the
desired fidelity order.  This section presents a simple generalization
of the usual error correction criteria which makes this idea
mathematically concrete. These {\em approximate error correction
criteria} are {\em sufficient} to do approximate error correction. We
expect that a more complete theory giving necessary and sufficient
conditions to do approximate error correction is possible, but have
not obtained such a theory. Nevertheless we hope that the
simple sufficient conditions presented here will inspire other
researchers to develop a general theory of approximate error
correction.

Quantum error correction is performed by encoding logical basis states
in a subspace ${\cal C}$ of the total Hilbert space ${\cal H}$ so that
some effects of ${\cal E}$ can be reversed on ${\cal C}$. 
Let the noise process be described in some {\em operator sum representation} 
${\cal E}(\rho) = \sum_{n \in {\cal K}} A_n \rho A_n^{\dagger}$, 
where ${\cal K}$ is the index set of ${\cal A}$, the set of all 
effects $A_n$ appearing in the sum.   
We denote by ${\cal A}_{re} \subset {\cal A}$ the reversible 
subset on ${\cal C}$, and let ${\cal K}_{re} = \{n|~A_n\in{\cal A}_{re}\}$ 
be the index set of ${\cal A}_{re}$.  In other words, the process   
${\cal E}'(\rho) = \sum_{n \in {\cal K}_{re}} A_n \rho A_n^{\dagger}$ 
is reversible on ${\cal C}$.  ${\cal A}_{re}$ includes all 
the effects satisfying the condition 
\cite{Ekert96b,Knill96,Bennett96a,Nielsen96c}
\begin{equation}
        P_C A_m^\dagger A_n P_C = g_{mn} P_C
        ~~~~~\forall m,n \in {\cal K}_{re}
\,,
\label{eq:pccondnon}
\end{equation}  
where $P_C$ is the projector onto ${\cal C}$, and $g_{mn}$ are
entries of a positive matrix.  When Eq.(\ref{eq:pccondnon}) is
true for some subset of effects $A_n$ which form an 
operator-sum representation for the operation ${\cal E}$, 
there exists some operator-sum representation of ${\cal E}$ such that 
${\cal E}'(\rho) = \sum_{n \in \tilde{{\cal K}}_{re}} 
\tilde{A}_n \rho \tilde{A}_n^{\dagger}$ and:
\begin{equation}
        P_C \tilde{A}_m^\dagger \tilde{A}_n P_C = p_{n} \delta_{mn} P_C
        ~~~~~\forall m,n \in \tilde{{\cal K}}_{re}
\,,
\label{eq:pccond}
\end{equation}  
where $p_n$ are non-negative c-numbers.
Eq.(\ref{eq:pccond}) (or equivalently Eq.(\ref{eq:pccondnon})) must 
be satisfied for some subset of effects $\tilde{A}_n$ ($A_n$) 
which form an operator-sum representation
for the operation ${\cal E}$.
We emphasize that the $A_n$ operators in Eq.(\ref{eq:pccondnon}) 
and the $\tilde{A}_n$  operators in Eq.(\ref{eq:pccond})
are not necessarily the same, and similar distinction holds for
the reversible subsets.  
For simplicity of notation, 
we will drop the tilde even when we refer to 
the operators and reversible subset in Eq.(\ref{eq:pccond}).

Eq.(\ref{eq:pccond}) is equivalent to the usual (exact) {\em
orthogonality} and {\em non-deformation} conditions for a {\em
non-degenerate} code with logical states $|c_i\rangle$, and these
conditions can be stated concisely as: $\forall i, j$,
\begin{equation}
        \langle  c_i | A^\dagger_m A_n |c_j \rangle
        = \delta_{ij}\delta_{mn} p_n  ~~~\forall m,n \in {\cal K}_{re}
\label{eq:critonetwo}
\end{equation} 
where $p_n$ are the non-negative c-numbers in Eq.(\ref{eq:pccondnon}), and
they represent the {\em error detection
probability} of the effects $A_n$.

When Eq.(\ref{eq:pccond}) is satisfied, the effects $A_n$ have polar
decompositions
\begin{equation}
        A_n P_C = \sqrt{p_n} U_n P_C
        ~~~~~\forall n\in {\cal K}_{re}
\,,
\label{eq:polardecom}
\end{equation} 
where the $U_n$'s are unitary.  Note that $P_C U_n^{\dagger} U_m P_C =
\delta_{nm} P_C$ is required for syndrome detection to be unambiguous.  
This is simple to see by writing the recovery operation ${\cal R}$ as
\begin{equation} 
	{\cal R}(\rho)=\sum_{k \in {\cal K}_{re}} R_k
        \rho R_k^{\dagger} + P_E \rho P_E
\label{eq:recov}
\,,
\end{equation}
where $R_k = P_C U_k^{\dagger}$ is the appropriate reversal process
for each effect $A_k$ ($k\in {\cal K}_{re}$), and 
$P_E \equiv I - \sum_{k \in {\cal K}_{re}} U_k P_C U_k^{\dagger}$. 
When we apply ${\cal R}$ on ${\cal E}(|\psi_{in}\rangle\langle\psi_{in}|)$,  
where $|\psi_{in}\rangle$ is a pure input state, the first term in 
Eq.(\ref{eq:recov}) is proportional to 
$|\psi_{in}\rangle\langle\psi_{in}|$, and represents the recovered state.
${\cal R}(\rho)$ is trace preserving as a result of  
orthogonality of different $U_k P_C$.

It is convenient to define $P^{det} = \sum_{n\in {\cal K}_{re}} p_n$
to be the total detection probability for the reversible subset.  
It is immediate that to achieve a desired
fidelity ${\cal F}$, we have to include in ${\cal A}_{re}$ a sufficient
number of highly probable effects so that $P^{det} \geq {\cal F}$.  
Thus, ${\cal A}_{re}$ is also a high probability subset.  

Now we generalize Eq.(\ref{eq:pccond})-(\ref{eq:polardecom}) based on the
following assumption: the error is parameterized by a certain number
of small quantities with physical origins such as the strength and
duration of the coupling between the system and the environment.  For
simplicity, we consider only one-parameter processes, and let
$\epsilon$ be the small parameter. For example, $\epsilon$ can be the
single qubit error probability. Suppose the aim is to find a code for
a known ${\cal E}$ with fidelity:
\begin{equation}
        {\cal F} \geq 1 - {\cal O}(\epsilon^{t+1})
\,.
\label{eq:fidcond}
\end{equation}
In the new criteria, it is still necessary that $P^{det} \geq {\cal F}$, 
that is, ${\cal A}_{re}$ has to include all effects 
$A_n$ with maximum detection probability 
$\max_{|\psi_{in}\rangle\in{\cal C}} \mbox{tr}
(|\psi_{in}\rangle\langle\psi_{in}| A_n^{\dagger} A_n) 
\approx {\cal O}(\epsilon^s)$, $s \leq t$.  
However, it is {\em not} necessary to recover the exact input state;
only a good overlap between the input and output states is needed.  In
terms of the condition on the codeword space, it suffices for the
effects to be {\em approximately} unitary and mutually orthogonal.
These observations can be expressed as relaxed {\em sufficient}
conditions for error correction. Suppose
\begin{eqnarray}
	A_n P_C = U_n \sqrt{P_C A_n^{\dagger} A_n P_C}
\label{eq:appolardecom}
\,, 
\end{eqnarray}
is a polar decomposition for $A_n$. We define c-numbers $p_n$ and
$\lambda_n$ so that $p_n$ and $p_n \lambda_n$ are the largest and the
smallest eigenvalues of $P_C A_n^{\dagger} A_n P_C$, considered as an
operator on ${\cal C}$. The relaxed conditions for error correction
are that:
\begin{eqnarray}
	p_n (1 - \lambda_n) 
        & \leq &  {\cal O}(\epsilon^{t+1})  ~~~~~\forall n \in {\cal K}_{re}
\label{eq:newcritone}
\\
        P_C U_m^{\dagger} U_n P_C & = & \delta_{mn} P_C
\label{eq:newcritthree}
\,.
\end{eqnarray} 
Note that when $\lambda_n=1$, Eq.(\ref{eq:appolardecom}) - 
(\ref{eq:newcritthree}) reduce to the exact criteria.  
In the approximate case, $P^{det} = \sum_{n \in {\cal K}_{re}} 
\mbox{tr}(|\psi_{in}\rangle\langle\psi_{in}| A_n^{\dagger} A_n)$ 
is not a constant, 
but depends on the input state $|\psi_{in}\rangle$.  
Since ${\cal A}_{re}$ includes enough effects so that
$P^{det} \geq 1-{\cal O}(\epsilon^{t+1})$, when 
Eq.(\ref{eq:newcritone}) is satisfied, we also have 
$\sum_{n \in {\cal K}_{re}} p_n \geq 1 - {\cal O}(\epsilon^{t+1})$ and  
$\sum_{n \in {\cal K}_{re}} p_n \lambda_n \geq 1 - {\cal O}(\epsilon^{t+1})$.

We now prove that $\sum_{n \in {\cal K}_{re}} p_n \lambda_n$ is 
a lower bound on
the fidelity. Defining the {\em residue operator}
\begin{eqnarray}
	\pi_n = \sqrt{P_C A_n^{\dagger} A_n P_C} - \sqrt{p_n \lambda_n} P_C
\,,
\end{eqnarray}
we find, for the operator norm of $\pi_n$, 
\begin{eqnarray}
%
	0 \leq |\pi_n| \leq \sqrt{p_n} - \sqrt{p_n \lambda_n}
\label{eq:positivity}
\,,
\end{eqnarray}
and
\begin{eqnarray} 
\label{eq: andecomp}
	A_n P_C = U_n( \sqrt{p_n \lambda_n} I + \pi_n )P_C
\,.
\end{eqnarray} 

The sufficiency of our conditions to obtain the desired fidelity may
be proved as follows.  Though Eq.(\ref{eq:polardecom}) is not satisfied,
as long as Eq.(\ref{eq:newcritthree}) is true, we 
can still define the {\em approximate} recovery operation
\begin{equation} 
	{\cal R}(\rho)=\sum_{k \in {\cal K}_{re}} R_k
        \rho R_k^{\dagger} + P_E \rho P_E
\label{eq:approxrec}
\end{equation}
with $R_k = P_C U_k^{\dagger}$ as the {\em approximate} recovery
operation for $A_k$ and $P_E$ defined as in the case of exact error 
correction.  For a pure input state $|\psi_{in}\rangle \langle \psi_{in}|$,
applying ${\cal R}(\rho)$ on ${\cal E}(|\psi_{in}\rangle \langle \psi_{in}|)$,
and ignoring the last term which is positive definite produces an output with
fidelity
\begin{eqnarray}
        {\cal F} ~ &\equiv& ~ \min_{|\psi_{in}\rangle \in {\cal C}} 
	\mbox{tr}(|\psi_{in}\rangle \langle \psi_{in}| 
		{\cal R}({\cal E}(|\psi_{in}\rangle \langle \psi_{in}|)))
\nonumber
\\
        &\geq&  \min_{|\psi_{in}\rangle \in {\cal C}} 
	\sum_{k,n \in {\cal K}_{re}} 
	|\langle \psi_{in}| U_k^{\dagger} A_n |\psi_{in} \rangle |^2
\,.
\end{eqnarray}
Omitting all terms for which $k \neq n$ and applying
Eq.(\ref{eq: andecomp}) gives
\begin{eqnarray}
\nonumber
	{\cal F} &\geq& \min_{|\psi_{in}\rangle \in {\cal C}}
	\sum_{n \in {\cal K}_{re}} 
	|\langle \psi_{in}| \sqrt{p_n \lambda_n} + \pi_n |\psi_{in}\rangle |^2 
\\	&\geq& \sum_{n \in {\cal K}_{re}} p_n \lambda_n 
\,,
\end{eqnarray}
where in the last step, we have used Eq.(\ref{eq:positivity}).  
Hence, the fidelity is at least 
$\sum_{n\in {\cal K}_{re}} p_n \lambda_n \geq 1 - {\cal O}(\epsilon^{t+1})$ 
and the desired fidelity order is achieved as claimed.

An explicit procedure for performing this recovery is as follows.
First, a measurement of the 
projectors $P_k \equiv U_k P_C U_k^{\dagger}$ is performed.
Conditional on the result $k$ of the measurement,
the unitary operator $U_k$ is applied to complete the recovery.

Note that when the exact criteria hold, it is not necessary for the set
${\cal A}$ to form an operator sum representation for the error process; 
they may instead give a generalized description of the quantum process 
such as
\begin{equation}
        {\cal E}(\rho) = \sum_{mn\in {\cal K}} A_m \rho A_n^{\dagger} \chi_{mn}
\,,
\end{equation}
where $\chi_{mn}$ is a matrix of c-numbers\cite{Chuang96d}.  In this
case, approximate criteria can be obtained straightforwardly along the
lines we have described for the operator sum representation (where
$\chi_{mn}$ is diagonal).  The main issue is to ensure that
interference terms coming from off-diagonal terms in $\chi_{mn}$ are
sufficiently small.  We will not describe this calculation here.
Instead, we return to our four-bit code, and analyze it using the new
criteria.
\section{4-bit Code Revisited}

In terms of the approximate quantum error correction criteria
Eqs.(\ref{eq:newcritone}-\ref{eq:newcritthree}), we may understand why
our four-bit amplitude damping quantum code works as follows.  Let the
orthonormal basis for the Hilbert space be ordered as:
\begin{equation}
        |0000\rangle , |0011\rangle , |1100\rangle , 
	|1111\rangle , |0111\rangle , |0100\rangle , \ldots
\end{equation}
We include only the first few non-zero rows and columns in our
matrices.  The projection operator 
$|0_L\rangle \langle 0_L|+|1_L\rangle \langle 1_L|$ onto
the two dimensional codeword space ${\cal C}$ is
\begin{equation}
        P_C = \frac{1}{2} {\left[
        \begin{array}{cccc}
            1 & 0 & 0 & 1 \\ 0 & 1 & 1 & 0
        \\  0 & 1 & 1 & 0 \\ 1 & 0 & 0 & 1
        \end{array} \right]}
\,
\end{equation}
The first effect operator (no loss of a quantum to the environment) is
\begin{equation}
        A_{0000} = {\left[
        \begin{array}{cccc}
            1 & 0 & 0 & 0 \\ 0 & 1-\gamma & 0 & 0
        \\  0 & 0 & 1-\gamma & 0 \\ 0 & 0 & 0 & (1-\gamma)^2
        \end{array} \right]}
\,.
\end{equation}
The eigenvalues of $P_C A_{0000}^{\dagger} A_{0000} P_C$ are
$(1-\gamma)^2$ and $\frac{1}{2}(1+(1-\gamma)^4)$.
Interested readers can check for themselves that
\begin{eqnarray}
        & & A_{0000} P_C
\nonumber
\\
        &=& U_{0000} \left[\rule{0pt}{2.4ex}{
	(1-\gamma) I + (\gamma^2 + {\cal O}(\gamma^4)) \tilde{\pi}_{0000} 
	}\right] P_C
\,
\end{eqnarray}
(the order of $\gamma$ in $\pi_{0000}$ is factored out of 
$\tilde{\pi}_{0000}$) with the choice:
\begin{eqnarray}
        U_{0000} &=& {\left[
        \begin{array}{cccc}
        \cos(\theta-\frac{\pi}{4}) & 0 & 0 & -\sin(\theta-\frac{\pi}{4})
\\
        0 & 1 & 0 & 0
\\
        0 & 0 & 1 & 0
\\
        \sin(\theta-\frac{\pi}{4}) & 0 & 0 & \cos(\theta-\frac{\pi}{4})
        \end{array} \right]}
\\
        \tilde{\pi}_{0000} &=& {\left[
        \begin{array}{cccc}
        1 & 0 & 0 & 0
\\
        0 & 0 & 0 & 0
\\
        0 & 0 & 0 & 0
\\
        0 & 0 & 0 & 1
        \end{array} \right]}
\,
\end{eqnarray}
where $\tan \theta = (1-\gamma)^2$.  The exact criteria are not
satisfied, as $P_C A^{\dagger}_{0000} A_{0000} P_C$ has different
eigenvalues.  However, the difference is of order ${\cal O}(\gamma^2)$
and thus the relaxed condition 
Eq.(\ref{eq:newcritone}) is satisfied.

For the second effect (loss of one quantum from the $n_1$ mode), we have
\begin{equation}
        A_{1000}  =  (1-\gamma)^{\frac{1}{2}}\sqrt{\gamma} {\left[
        \begin{array}{cccccc}
            0 & 0 & 0 & 0 & 0 & 0
\\
            0 & 0 & 0 & 0 & 0 & 0
\\
            0 & 0 & 0 & 0 & 0 & 0
\\
            0 & 0 & 0 & 0 & 0 & 0
\\
            0 & 0 & 0 & 1-\gamma & 0 & 0
\\
            0 & 0 & 1 & 0 & 0 & 0
        \end{array} \right]}
\,.
\end{equation}
The eigenvalues of $P_C A^{\dagger}_{1000} A_{1000} P_C$ are $\gamma
(1-\gamma)$ and $\gamma (1-\gamma)^3$.  The difference is $(2 \gamma^2
- \gamma^3) (1 - \gamma)$.  We have the decomposition
\begin{eqnarray}
        & & A_{1000} P_C
\nonumber
\\
        &=& \sqrt{\frac{(1-\gamma) \gamma} {2}} U_{1000} 
	\left[\rule{0pt}{2.4ex}{
	(1-\gamma) I + \gamma \tilde{\pi}_{1000} }\right] P_C
\,
\end{eqnarray}
\begin{eqnarray}
        U_{1000} & = & {\left[
        \begin{array}{cccccc}
            1 & 0 & 0 & 0 & 0 & 0
\\
            0 & 1 & 0 & 0 & 0 & 0
\\
            0 & 0 & 0 & 0 & 0 & 1
\\
            0 & 0 & 0 & 0 & 1 & 0
\\
            0 & 0 & 0 & 1 & 0 & 0
\\
            0 & 0 & 1 & 0 & 0 & 0
        \end{array} \right]}
        {\left[
        \begin{array}{cccccc}
            \frac{1}{\sqrt{2}} & 0 & 0 & -\frac{1}{\sqrt{2}} & 0 & 0
\\
            0 & \frac{1}{\sqrt{2}} & -\frac{1}{\sqrt{2}} & 0 & 0 & 0
\\
            0 & \frac{1}{\sqrt{2}} & \frac{1}{\sqrt{2}} & 0 & 0 & 0
\\
            \frac{1}{\sqrt{2}} & 0 & 0 & \frac{1}{\sqrt{2}} & 0 & 0
\\
            0 & 0 & 0 & 0 & 1 & 0
\\
            0 & 0 & 0 & 0 & 0 & 1
        \end{array} \right]}
\nonumber
\\
        \tilde{\pi}_{1000} & = &  {\left[
        \begin{array}{cccccc}
            0 & 0 & 0 & 0 & 0 & 0
\\
            0 & 0 & \frac{1}{\sqrt{2}} & 0 & 0 & 0
\\
            0 & 0 & \frac{1}{\sqrt{2}} & 0 & 0 & 0
\\
            0 & 0 & 0 & 0 & 0 & 0
\\
            0 & 0 & 0 & 0 & 0 & 0
\\
            0 & 0 & 0 & 0 & 0 & 1
        \end{array} \right]}
\,.
\end{eqnarray}
Other one loss cases are similar. The $U_k P_C$ have non-zero entries
in different rows, and are orthogonal to each other.  Hence, all the
approximate code criteria are satisfied.  Using these explicit matrices,
we obtain
\begin{eqnarray}
        {\cal R}({\cal E}(|\psi_{in}\rangle \langle \psi_{in}|))
        & \approx & \sum_{k \in {\cal K}_{re}}P_C U_k^{\dagger} A_k 
	|\psi_{in}\rangle \langle \psi_{in}| A_k^{\dagger} U_k P_C
\nonumber
\\
        &=& (1-3 \gamma^2) |\psi_{in}\rangle \langle \psi_{in}| + \ldots
\,.
\label{eq:recstate}
\end{eqnarray}
The fidelity is thus at least $1-3 \gamma^2$, and is of the desired order.

The recovery procedure suggested in Eq.(\ref{eq:recstate}) contrasts 
with the decoding and recovery circuits in section \ref{sec:reco}.  
It is an interesting exercise to check that the composition of the 
operations in Fig.~\ref{fig:fbitall}A and~\ref{fig:fbitall}B, 
followed by re-encoding the recovered qubit has the same effect 
on ${\cal C}$ as applying $U_{0000}^{\dagger}$ for recovery.  
For the case in which an emission occurs in the first qubit, the
composition of operations in Fig.~\ref{fig:fbitall}A and~\ref{fig:fbitall}C, 
followed by re-encoding the recovered qubit has the same effect 
on ${\cal C}$ as preferentially damp out the $|n_3\rangle = |0\rangle$ 
component followed by applying $U_{1000}^{\dagger}$ for recovery. 
Note that it costs
$2 \gamma^2$ in the fidelity for removing the distortion.

\section{Applications to other codes}

Our approximate criteria may also be used to simplify code construction 
using a non-Pauli error description basis.  For
example, consider the bosonic quantum codes for amplitude
damping\cite{Chuang96c} (these are codes which utilize bosonic states
$|0\rangle , \cdots |n\rangle $ instead of qubits).  For logical states
$|c_1\rangle ,|c_2\rangle ,\ldots$ of the form:
\begin{eqnarray}
        |c_l\rangle &=&\sqrt{\mu_1}|n_{11} n_{12} \ldots n_{1m}\rangle 
\nonumber
\\      
      	&+&\sqrt{\mu_2}|n_{21} n_{22} \ldots n_{2m}\rangle 
\nonumber
\\            
	&+&\cdots
\nonumber
\\            
	&+&\sqrt{\mu_{N_{l}}}|n_{N_{l}1} n_{N_{l}2} \ldots n_{N_{l}m}\rangle 
\,,
\end{eqnarray}
the original non-deformation conditions for correcting up to one photon
loss require the following to be constant for all logical states:
\begin{eqnarray}
        \langle c_l|A_0^{\dagger}A_0|c_l\rangle 
        &=&\sum_{i=1}^{N_l}(1-\gamma)^{RS_i}\mu_i
\\      
\langle c_l|A_{0\cdots1\cdots0}^{\dagger}A_{0\cdots1\cdots0}|c_l\rangle 
        &=&\sum_{i=1}^{N_l}(1-\gamma)^{RS_i-1} \gamma \mu_i n_{ij}
\,.
\label{eq:c0}
\end{eqnarray} 
In the above, $RS_i=\sum_{j=1}^{m}n_{ij}$ is the total number of
excitation in the $i^{th}$ quasi-classical state (QCS) in $|c_l\rangle$.
It is difficult to find a solution for Eq.(\ref{eq:c0}) 
when $RS_i$ is not constant for
all $i$.  That is the reason why previous work
\cite{Plenio96a,Chuang96a} suggests using only QCS with the same
number of excitations.

With the new criteria, it suffices for the following to be constant for 
all logical states:
\begin{equation}
\label{eq:c1}
        \sum_{i=1}^{N_l}\gamma \mu_i n_{ij} ~~~~~~\forall j
\,.
\end{equation}
That is, instead of requiring the individual number of excitations in
each QCS to be balanced, it is sufficient to balance the {\em average}
number of excitations over the QCS in each codeword.  
This provides an alternative explanation
of why the known five-bit perfect quantum codes \cite{Bennett96a,Laflamme96}
\begin{eqnarray}
        |0_L\rangle 
	&=&|00000\rangle +|11000\rangle -|10011\rangle -|01111\rangle 
\nonumber
\\            
	&+&|11010\rangle +|00110\rangle +|01101\rangle +|10101\rangle 
\nonumber
\\      
	|1_L\rangle 
	&=&|11111\rangle -|00011\rangle +|01100\rangle -|10000\rangle 
\nonumber
\\            
	&-&|00101\rangle +|11010\rangle +|10010\rangle -|01010\rangle 
\end{eqnarray}
work for amplitude damping errors (as described in Eq.(\ref{eq:ampdamperr})):
although the codes do not satisfy the exact non-deformation criterion,
Eq.(\ref{eq:c0}) for the error representation of
Eqs.(\ref{eq:adzero}-\ref{eq:adone}), they do satisfy the {\em approximate}
ones leading to Eq.(\ref{eq:c1}).

\section{Conclusion}

\label{sec:final}

We have shown by an example that choosing an appropriate error basis can
potentially reduce the number of qubits and other requirements in coding
schemes.  We also suggested a method to enable code construction without the
Pauli basis to be done more easily.  We believe that much more can be done
along these two directions.  The new sufficient criteria for error correction
are far from being necessary.
Our results show that it is worthwhile to look for better approximate
conditions or conditions which are both necessary and sufficient.  Approximate
error correction is particularly interesting because it is a property with no
classical analogue, as it makes use of slight non-orthogonalities possible
only between quantum states.  It would be especially useful to develop a
general framework for constructing codes based on approximate conditions,
similar to the group-theoretic framework now used to construct codes which
satisfy the exact conditions.

\acknowledgements

We thank David~DiVincenzo and John~Preskill, for useful and enjoyable
discussions about quantum error-correction. MAN would especially like
to thank Howard Barnum and Carlton Caves for explaining the
error-correction conditions to him.  This work was supported in part
by the Office of Naval Research (N00014-93-1-0116). We thank the
Institute for Theoretical Physics for its hospitality and for the
support of the National Science Foundation (PHY94-07194). MAN
acknowledges financial support from the Australian-American
Educational Foundation (Fulbright Commission).  DWL was supported in
part by the Army Research Office under grant no.  DAAH04-96-1-0299.


\end{document}